\documentclass[iop,apj]{emulateapj}
\usepackage{apjfonts}

\begin{document}

\title{GRB 090510: a disguised short GRB with the highest Lorentz factor and circumburst medium.}

\shorttitle{GRB 090510}

\shortauthors{M. Muccino et al.}

\author{M. Muccino\altaffilmark{1,2}, R. Ruffini\altaffilmark{1,2,3}, C.L. Bianco\altaffilmark{1,2}, L. Izzo\altaffilmark{1,2}, A.V. Penacchioni\altaffilmark{1,3}, G.B. Pisani\altaffilmark{1,3}}

\altaffiltext{1}{Dip. di Fisica and ICRA, Sapienza Universit\`a di Roma, Piazzale Aldo Moro 5, I-00185 Rome, Italy.}
\altaffiltext{2}{ICRANet, Piazza della Repubblica 10, I-65122 Pescara, Italy.}
\altaffiltext{3}{Universite de Nice Sophia Antipolis, Nice, CEDEX 2, Grand Chateau Parc Valrose.}
 
\begin{abstract}
GRB 090510, observed both by Fermi and AGILE satellites, is the first bright short-hard Gamma-Ray Burst (GRB) with an emission from the keV up to the GeV energy range.
Within the Fireshell model, we interpret the faint precursor in the light curve as the emission at the transparency of the expanding $e^+e^-$ plasma: the Proper-GRB (P-GRB). 
From the observed isotropic energy we assume a total plasma energy $E^{tot}_{e^+e^-}=(1.10\pm0.06)\times10^{53}$erg and derive a Baryon load $B=(1.45\pm0.28)\times10^{-3}$ and a Lorentz factor at transparency $\Gamma_{tr}=(6.7\pm1.6)\times10^2$.
The main emission $\sim0.4$s after the initial spike is interpreted as the extended afterglow, due to the interaction of the ultrarelativistic baryons with the CircumBurst Medium (CBM). 
Using the condition of fully radiative regime, we infer a CBM average spherically symmetric density of $\langle n_{CBM}\rangle=(1.85\pm0.14)\times10^3$ particles/cm$^3$, one of the highest found in the Fireshell model.
The value of the filling factor, $1.5\times10^{-10}\leq\mathcal{R}\leq3.8\times10^{-8}$, leads to the estimate of filaments with densities $n_{fil}=n_{CBM}/\mathcal{R}\approx(10^{6}-10^{14})$ particles/cm$^3$.
The sub-MeV and the MeV emissions are well reproduced.
When compared to the canonical GRBs with $\langle n_{CBM} \rangle\approx1$ particles/cm$^3$ and to the disguised short GRBs with $\langle n_{CBM}\rangle\approx10^{-3}$ particles/cm$^3$, the case of GRB 090510 leads to the existence of a new family of bursts exploding in an over-dense galactic region with $\langle n_{CBM}\rangle\approx10^3$ particles/cm$^3$.
The joint effect of the high $\Gamma_{tr}$ and the high density compresses in time and ``inflates'' in intensity the extended afterglow, making it appear as a short burst, which we here define as ``disguised short GRB by excess''.
The determination of the above parameters values may represent an important step towards the explanation of the GeV emission.
\end{abstract}


\section{Introduction}

\begin{figure*}
\centering
\hfill
\includegraphics[width=0.45\hsize,clip]{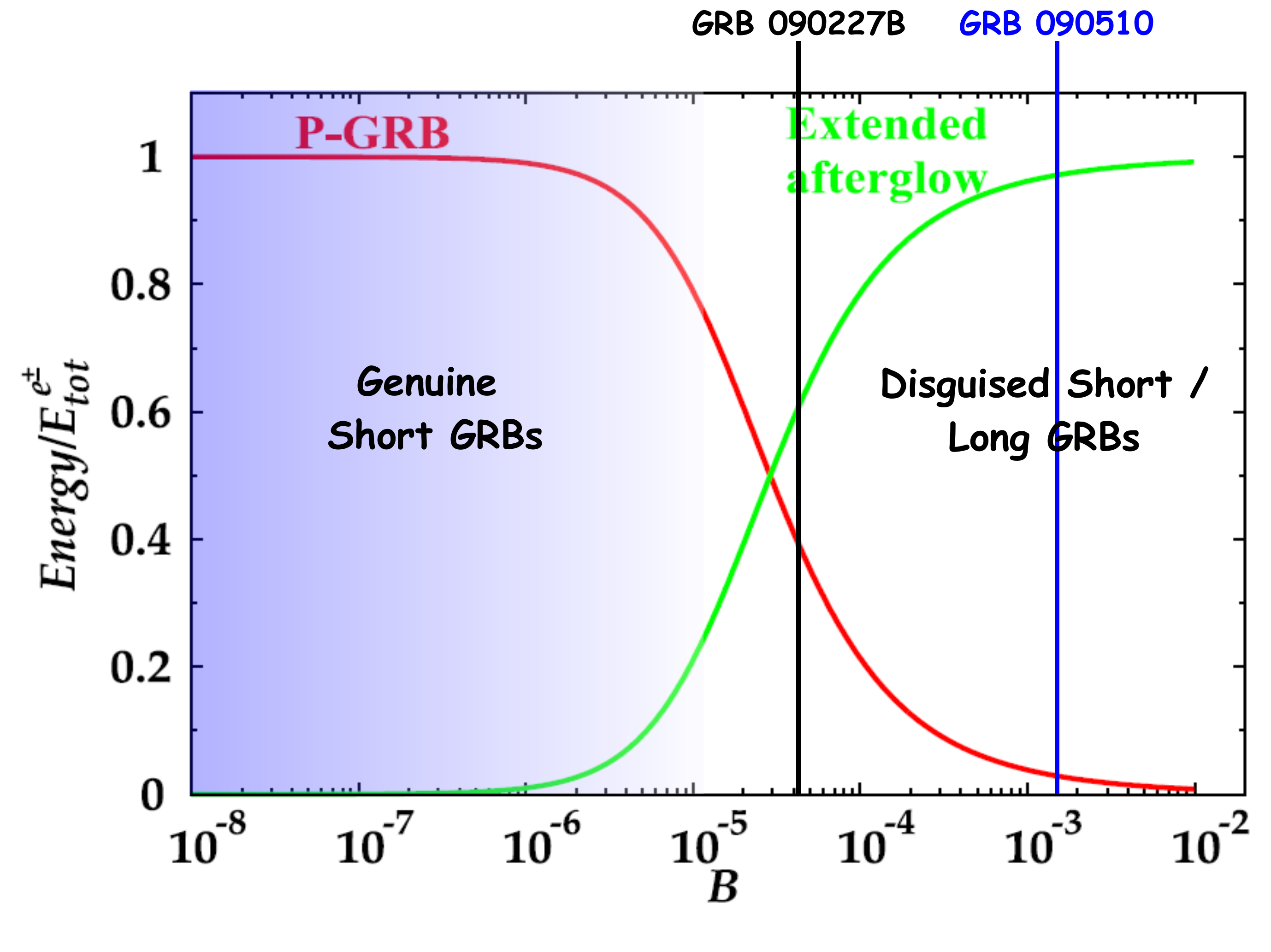}
\hfill
\includegraphics[width=0.45\hsize,clip]{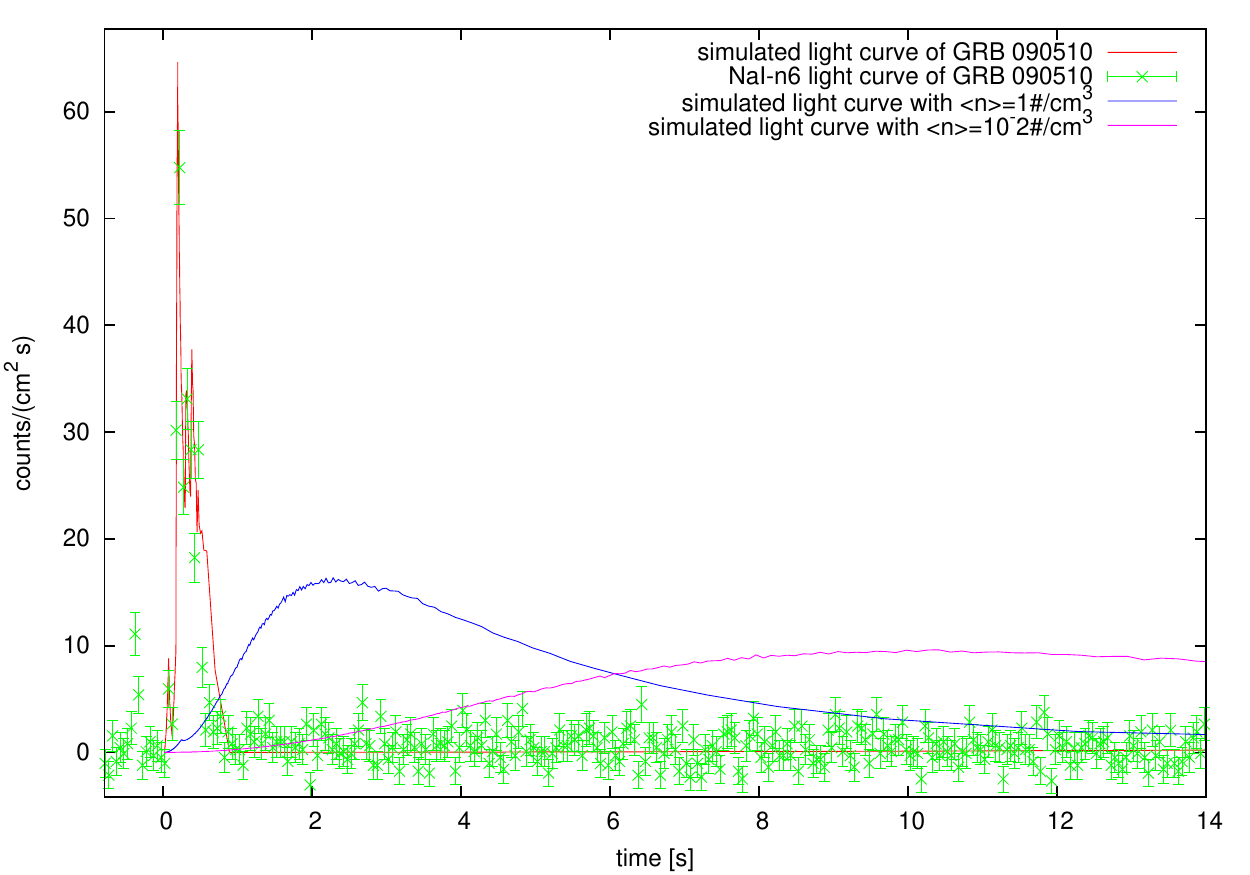}
\hfill\null
\caption{Upper panel: the energy emitted in the extended afterglow (green curve) and in the P-GRB (red curve) in units of $E_{e^+e^-}^{tot}$ are plotted as functions of $B$. The values of $B$ of GRB 090510 (in blue) and of the genuinely short GRB 090227B (in black) are compared. Lower panel: the $50$ ms time-binned NaI-n6 light curve (green data) and the extended afterglow simulations corresponding to CBM average densities of a ``disguised short GRB by excess'' with $\langle n_{CBM} \rangle\approx10^3$ particles/cm$^3$ (red curve), of a canonical long GRB with $\langle n_{CBM} \rangle=1$ particle/cm$^3$ (blue curve), and of a ``disguised short GRB by defect'' with $\langle n_{CBM} \rangle=10^{-2}$ particles/cm$^3$ (purple curve). For larger densities the extended afterglow compresses in time and ``inflates'' in intensity.}
\label{fig:1}
\end{figure*}

In their earliest classification of the $4$BATSE catalog \citep{Meegan1997}, all GRBs have been classified in short and long bursts being their $T_{90}$ duration longer or shorter than $2$ s  \citep{Klebesadel1992,Dezalay1992,Koveliotou1993,Tavani1998}.
In the meantime, short bursts have been shown to originate from a variety of astrophysical origins and not form a homogeneous class.
In the Fireshell model \citep{Ruffini2001c,Ruffini2001,Ruffini2001a,RVX2010}, the canonical GRB has two components: an emission occurring at the transparency of the optically thick expanding $e^+e^-$-baryon plasma \citep{RSWX}, the Proper-GRB (P-GRB), followed by the extended afterglow, due to the interactions between the accelerated baryons and the CircumBurst Medium (CBM). Such an extended afterglow comprises the prompt emission as well as the late phase of the afterglow \citep{Bianco2005b,Bianco2005a}.
The relative energy of these two components, for a given total energy of the plasma $E_{e^+e^-}^{tot}$, is uniquely a function of the baryon load $B=M_Bc^2/E^{tot}_{e^+e^-}$, where $M_B$ is the total baryons mass (see Fig.~\ref{fig:1}, upper panel).

The genuine short GRBs \citep{Ruffini2001} are the bursts occurring for $B\lesssim10^{-5}$.
The first example of such systems has indeed been recently identified, originating in a binary neutron star merger \citep{Muccino2012}.

It has been also proved the existence of disguised short GRBs, with baryon load $3\times10^{-4}\leq B\leq10^{-2}$ \citep{Bernardini2007,Bernardini2008}. 
In this class the extended afterglow is indeed energetically predominant but results in a ``deflated'' emission, less intense than the P-GRB, due to the low density of the CBM, $\langle n_{CBM} \rangle\approx10^{-3}$ particle/cm$^3$, much lower than the canonical value $\langle n_{CBM} \rangle\approx1$ particle/cm$^3$.
The majority of the declared short bursts in the current literature appears to be disguised short GRBs \citep{Bernardini2007,Bernardini2008,Caito2009,Caito2010,deBarros2011}.

In this paper we show a yet different kind of a disguised short, GRB 090510, again, with $3\times10^{-4}\leq B\leq10^{-2}$ and Lorentz factor $\Gamma_{tr}\approx700$, occurring in a medium with $\langle n_{CBM} \rangle\approx10^3$ particles/cm$^3$.
We define, indeed, these GRBs as ``disguised short burst by excess'', being their $\langle n_{CBM} \rangle$ much larger than the canonical one.
Correspondingly, we indicate the disguised short with a CBM density typical of the galactic halo environments, $\langle n_{CBM} \rangle\approx10^{-3}$ particles/cm$^3$, as ``disguised short GRBs by defect'' (see Fig.~\ref{fig:1}, lower panel).

The possibility of GRBs exploding in high density CBM has been already considered in literature \citep{Dai1999,Lazzati1999,Piro2001,Wang2003,Prochaska2008,Izzo2012b}.
In \citet{Dai1999,Piro2001,Wang2003}, the high density has been inferred from the steepening in the afterglows, respectively, of GRB 990123 in the $R$-band about $\sim2.5$ days after the burst, of GRB 000926 in the $R$-band after $\sim2$ days, and of GRB 990705 in the $H$-band after $\sim1$ day, due to the transition to the nonrelativistic regime of the fireball.
\citet{Lazzati1999} discuss the possibility that the detection of Fe lines in the afterglows of GRB 970508 and GRB 970828 could be due to recombination processes in extremely high densities during the X-ray afterglow.
In \citet{Prochaska2008}, the authors inferred dense environments, $n\gtrsim10^3$ particles/cm$^3$, from a survey for \ion{N}{5} absorption in GRBs afterglow spectra.
In particular, in \citet{Izzo2012b} the Fireshell model has been applied in the analysis of GRB 970828, discussed also in \citet{Lazzati1999}, inferring a dense environment with $\langle n_{CBM}\rangle=3.4\times10^3$ particle/cm$^3$, consistent with the large column density environment in \citet{Yoshida2001}.
In the case of GRB 090510 the joint effect of the very dense CBM and the high Lorentz factor at the transparency, $\Gamma_{tr}\sim700$, leads to an extended afterglow with $T_{90}<2$ s (see Fig.~\ref{fig:1}, lower panel). 
These high values of the CBM density $n_{CBM}$ and of the Lorentz factor $\Gamma_{tr}$ may represent an important step towards the explanation of the GeV emission.

The work is organized as follow: in Sec.~\ref{sec2} we present the data analysis of GRB 090510; in Sec.~\ref{sec3} we give our theoretical interpretation on the source; in Sec.~\ref{sec4} we summarize our conclusions.
 
\section{GRB 090510 Data Analysis}\label{sec2}

At 00:22:59.97 UT on 10$^{th}$ May 2009, the Fermi-GBM detector \citep{GCN9336} triggered and located the short and bright burst, GRB 090510, which was also detected by Swift \citep{GCN9331}, Fermi-LAT \citep{GCN9334}, AGILE \citep{GCN9343}, Konus-Wind \citep{GCN9344}, and Suzaku-WAM \citep{GCN9355}.
Optical observations by VLT/FORS2 located the host galaxy of GRB 090510 at the redshift of $z=0.903\pm0.003$ \citep{GCN9353}.
The offset with respect to the Nordic Optical Telescope refined afterglow position \citep{GCN9338} corresponds to $5.5$ kpc.

We have analyzed the Fermi-GBM data from NaI-n6 ($8$ -- $900$ keV) and BGO-b1 ($260$ keV -- $40$ MeV) detectors and the LAT data in the energy range $100$ MeV -- $30$ GeV.

\begin{figure*}
\centering
\hfill
\includegraphics[width=0.45\hsize,clip]{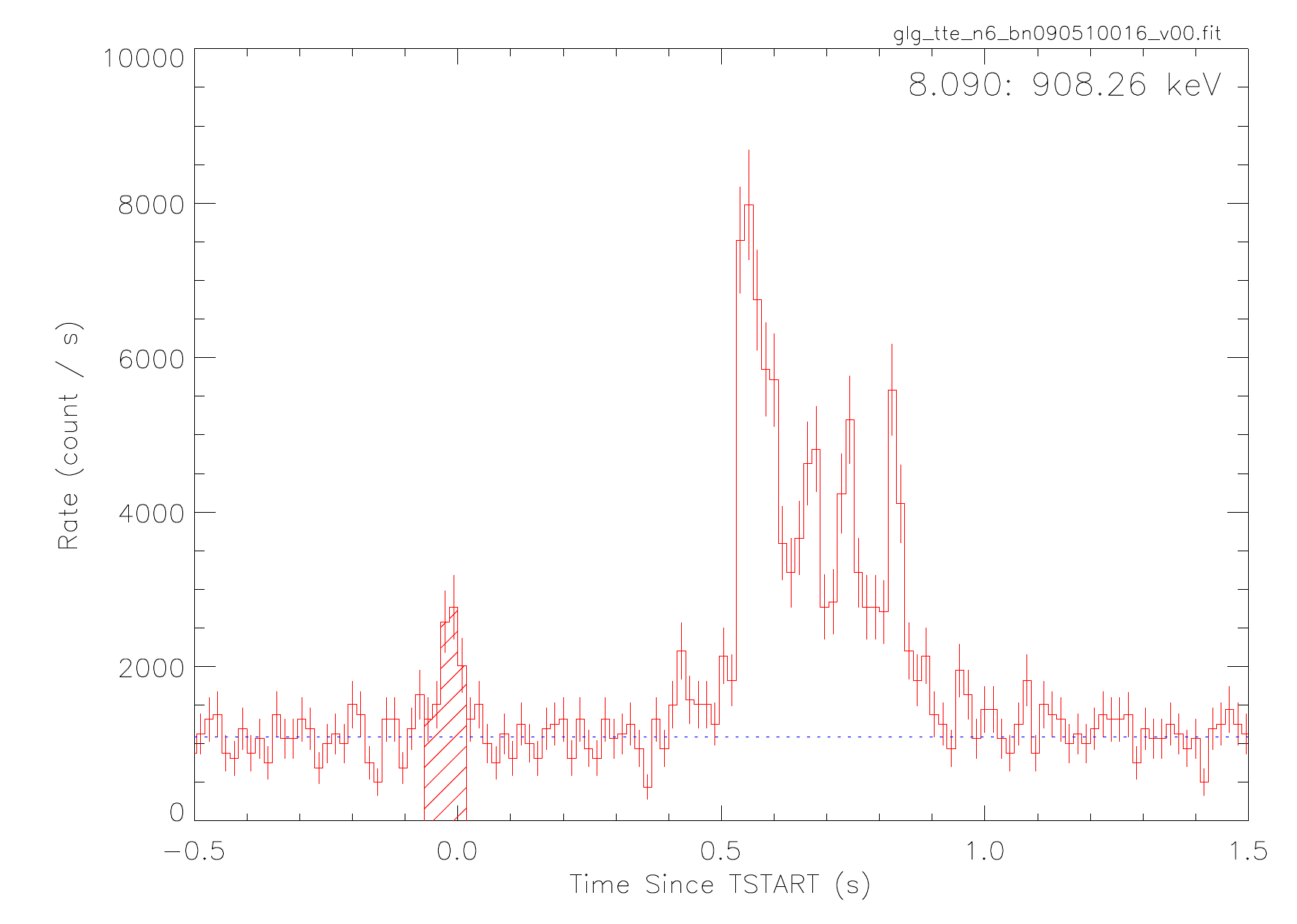}
\hfill
\includegraphics[width=0.45\hsize,clip]{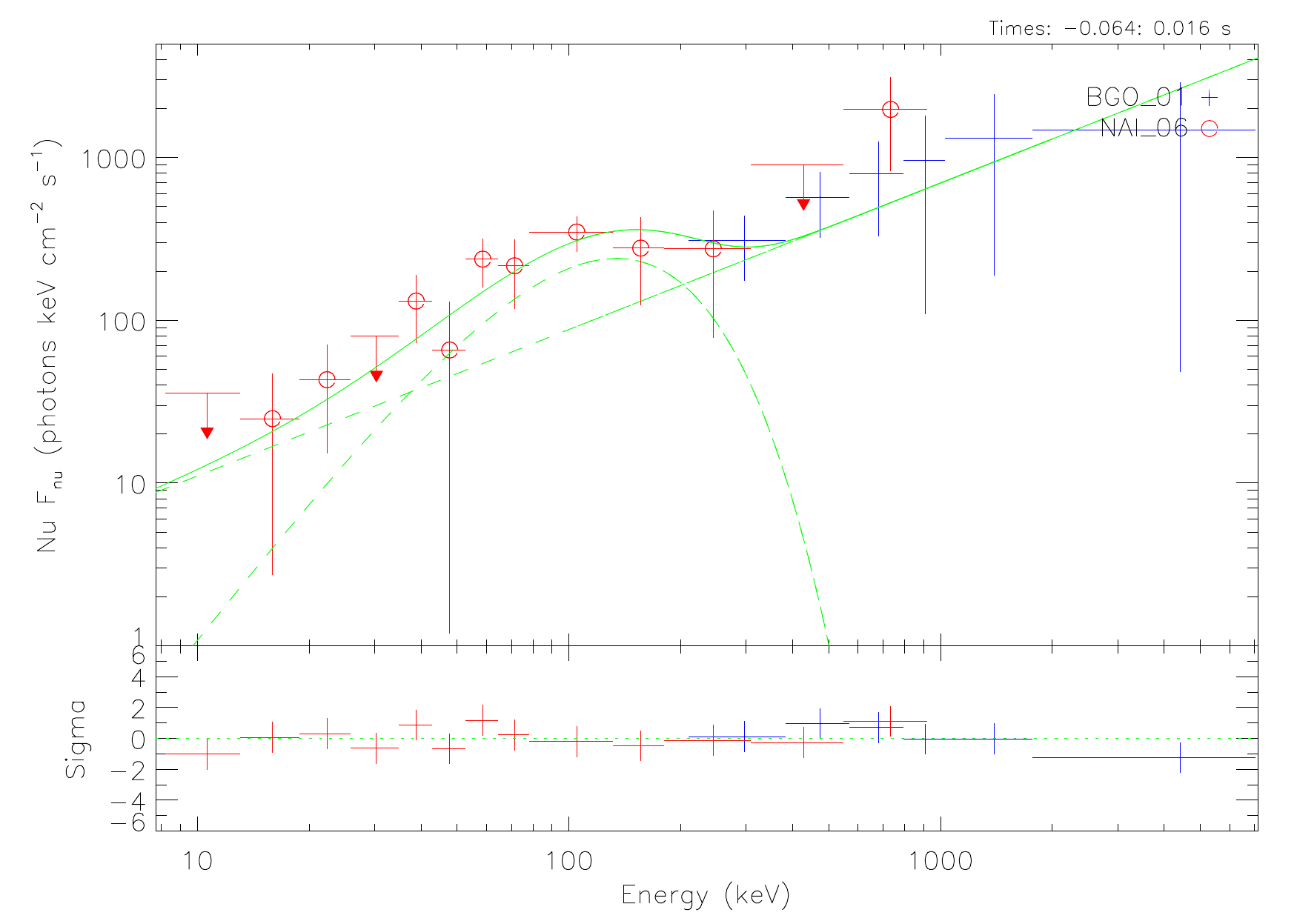}
\hfill\null\\
\hfill
\includegraphics[width=0.45\hsize,clip]{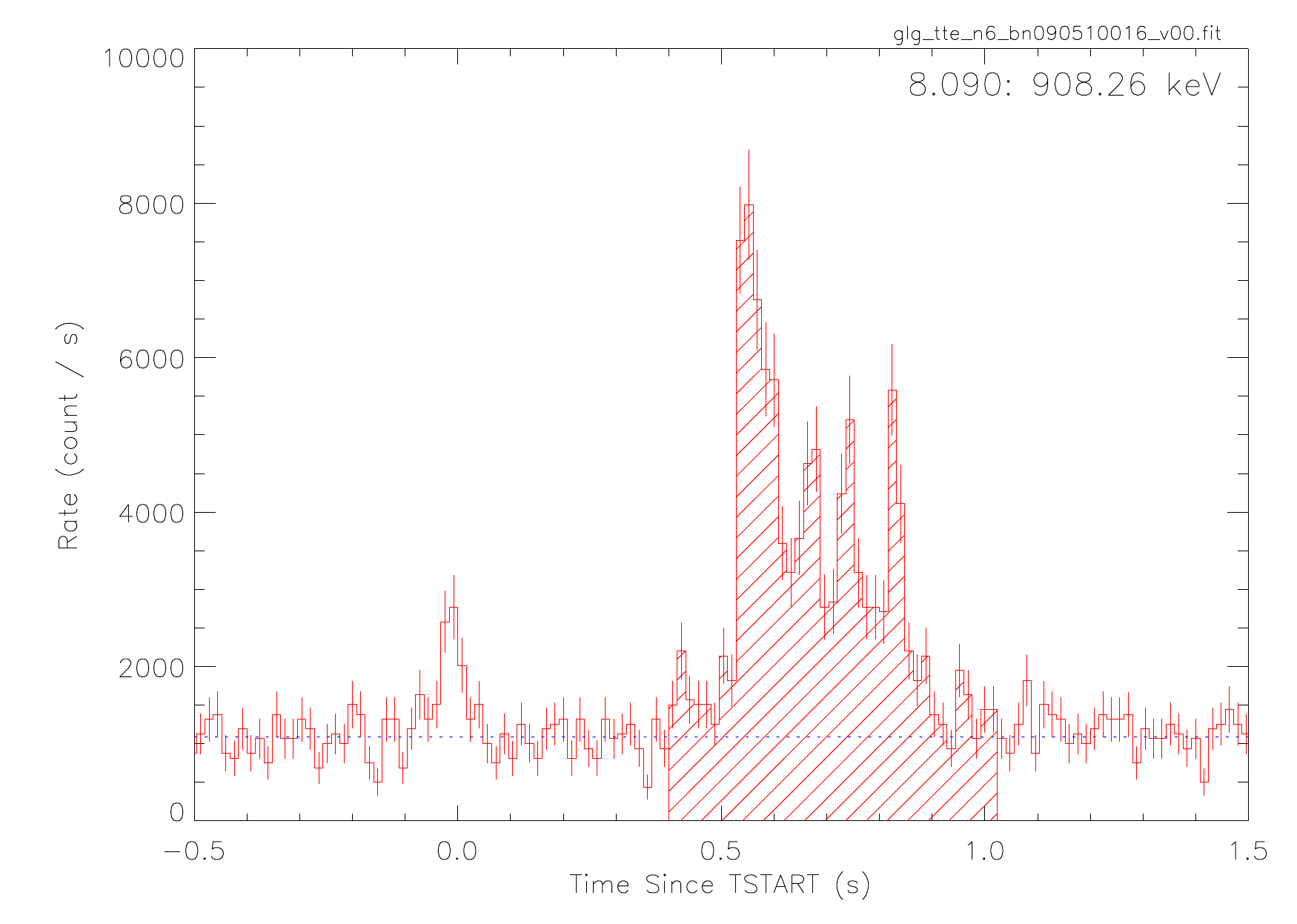}
\hfill
\includegraphics[width=0.45\hsize,clip]{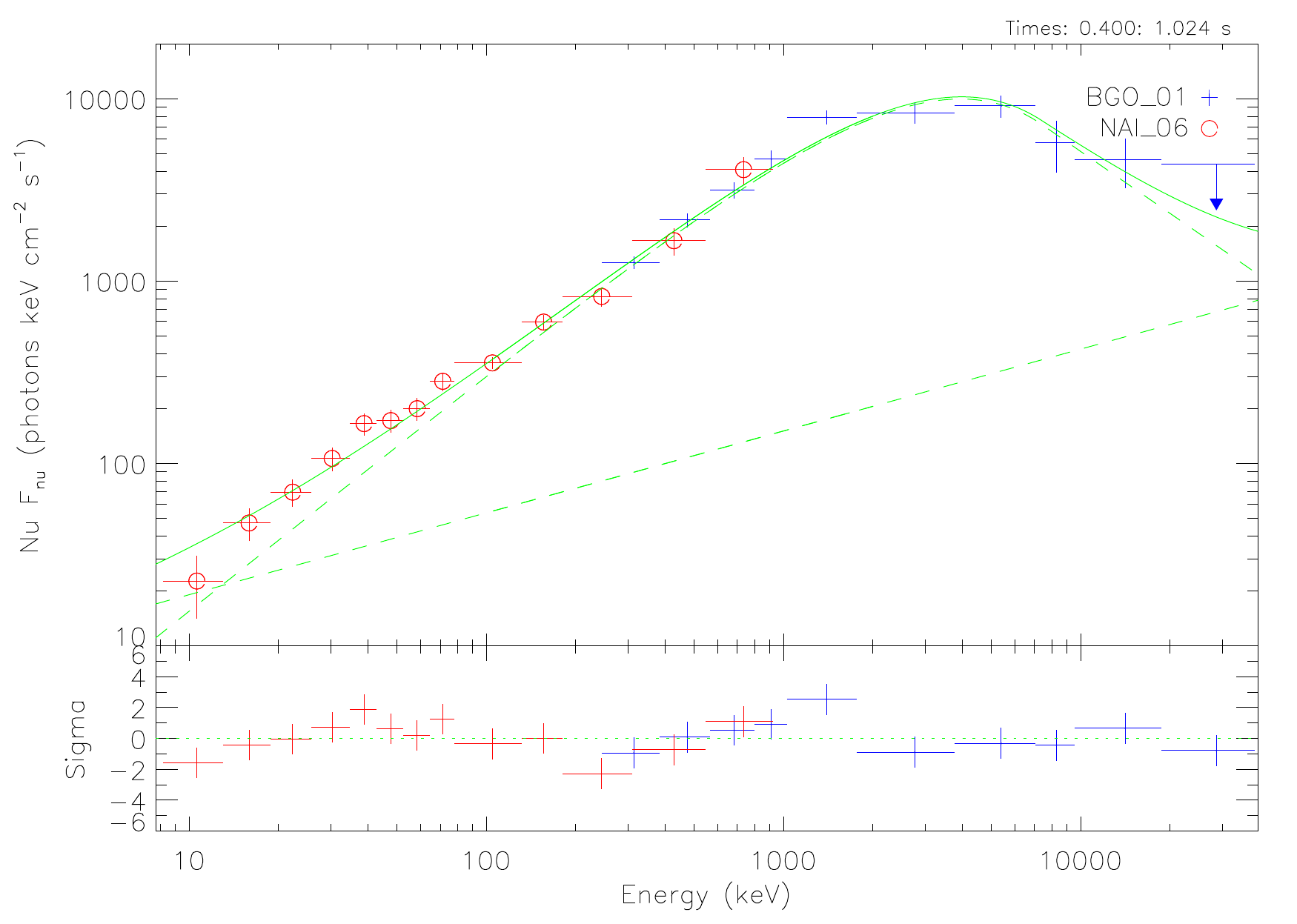}
\hfill\null\\
\caption{Upper panels: on the left, the 16 ms time-binned NaI-n6 light curve and, on the right, the NaI-n6+BGO-b1 $\nu F_\nu$ spectrum (best fit BB+PL) in the $\Delta T_1$ time interval. Lower panels: on the left, the 16 ms time-binned NaI-n6 light curve and, on the right, the NaI-n6+BGO-b1 $\nu F_\nu$ spectrum (best fit Band+PL) in the $\Delta T_2$ time interval.}
\label{fig:2}
\end{figure*}

The light curve of GRB 090510 is composed of two different episodes, $0.5$ s apart.
The first episode, from $T_0-0.064$ s to $T_0+0.016$ s (in the following $\Delta T_1$; $T_0$ is the trigger time), has not been considered by \citet{Ackermann2010}, \citet{Giuliani2010} and \citet{Guiriec2010} because of the small content of detected photons.
Even though the statistical content of this first episode is very poor, in this letter we show its great relevance for the theoretical analysis, since it can be identified with the P-GRB.
The second episode can be interpreted as the extended afterglow.
In the statistical analysis of the first episode, we have considered power-law (PL), black body (BB) plus PL, Band \citep{Band1993}, Comptonized (Compt), Band+BB and Compt+BB models.
Following the statistical analysis for nested models by \citet{Guiriec2010}, models more complicated than the simplest Band and Compt are singled out (see the last column of Tab.~\ref{sign}).
The direct statistical comparison between BB+PL and PL models gives a significance level of $3\%$ (see Tab.~\ref{sign}). 
This means that the BB+PL model improves the fit of the data of the first episode with respect to the PL model, which is excluded at $97\%$ confidence level.
The simple Band model has an unconstrained $\alpha$ index and a large error on the energy peak $E_p$, as well as in the case of the Compt model, for which the total flux is underestimated with respect to the Band and BB+PL models.
The quality of data does not allow us to favor the BB+PL model versus the Compt one from a pure statistical analysis. 
In order to clarify such a fundamental issue, it is appropriate that future space missions with larger collecting area and X/$\gamma$-rays timing be flown in the near future \citep[see e.g. LOFT mission,][]{LOFT2012}.
From our theoretical interpretation the BB+PL, being equally probable than the Compt model, is adopted for its physical meaning and because it is not ruled out by the data.
The BB observed temperature is $kT_{obs}=(34.2\pm7.5)$ keV (see Fig.~\ref{fig:2}, top right panel, and table below) and the total energy of the first episode is $E_1=(2.28\pm0.39)\times10^{51}$ erg.

We have then analyzed the second episode in the time interval from $T_0+0.400$ s to $T_0+1.024$ s (in the following $\Delta T_2$).
The best fit in the energy range $8$ keV -- $40$ MeV is Band+PL \citep{Ackermann2010} or, alternatively Compt+PL \citep{Giuliani2010,Guiriec2010}. 
The results are shown in Fig.~\ref{fig:2} and in Tab.~\ref{sign}.
Including the LAT data, the spectrum is again best fitted by Band+PL (see last row in Tab.~\ref{sign}), with the PL observed up to $30$ GeV \citep{Ackermann2010}. 
The total energy is $E_2=(1.08\pm0.06)\times10^{53}$ erg.
\begin{table*}
\centering
\begin{tabular}{ccccccccccc}
\hline\hline
\textbf{Interval}  &  \textbf{Model}        &  $kT$            & $\alpha$           &  $\beta$         &  $E_p$           &  $\gamma$            &  $F_{tot}\times10^{-6}$                         & C-STAT/DOF     &  \textbf{Significance}  \\
                   &                        &  [keV]           &                    &                  &  [keV]           &                      &  [erg/(cm${}^2$s)]                              &            \\
\hline
                   &  \textbf{PL}           &  $...$           &  $...$             &  $...$           &  $...$           &  $-1.22 \pm 0.06$    &  $9.2 \pm 1.3$                                  &  $195.41/195$  &                         \\
                   &  \textbf{BB+PL}        &  $34.2 \pm 7.5$  &  $...$             &  $...$           &  $...$           &  $-1.10 \pm 0.14$    &  $7.6 \pm 1.3$                                  &  $188.60/193$  &   $0.03$                \\
                   &  \textbf{Band}         &  $...$           &  unc               &  $-1.44\pm0.11$  &  $94\pm74$       &  $...$               &  $7.4 \pm 1.5$                                  &  $187.11/193$  &                         \\
$\Delta T_1$       & \textbf{Compt}         &  $...$           &  $-0.81\pm0.22$    &  $...$           &  $990\pm554$     &  $...$               &  $4.4 \pm 1.6$                                  &  $189.97/194$  &                         \\
                   &  \textbf{Band+BB}      &  $24.3 \pm 5.6$  &  unc               &  $-1.76\pm0.62$  &  unc             &  $...$               &  $7.1 \pm 2.0$                                  &  $186.90/191$  &  $0.57$                 \\
                   &  \textbf{Compt+BB}     &  $27.2 \pm 6.7$  &  $-0.72\pm0.39$    &  $...$           &  $2967\pm1570$   &  $...$               &  $8.4 \pm 2.3$                                  &  $187.23/192$  &  $0.90$                 \\
\hline
$\Delta T_2$ (a)   &  \textbf{Band+PL}      &  ...             &  $-0.70 \pm 0.10$  &  $-3.13\pm0.97$   &  $3941\pm 346$  &  $-1.55 \pm 0.54$    &  $43.6\pm 1.9$                                  &  $207.78/236$  &                         \\
\hline
$\Delta T_2$ (b)   &  \textbf{Band+PL}      &  ...             &  $-0.71 \pm 0.07$  &  $-2.97\pm0.26$   &  $4145\pm 398$  &  $-1.62 \pm 0.05$    &  $83.3\pm 6.8$                                  &  $199.20/256$  &                         \\
\hline
\end{tabular}
\caption{$\Delta T_1$ time interval: parameters of PL, BB+PL, Band, Compt, Band+BB and Compt+BB models in the energy range $8$--$7000$ keV. $\Delta T_2$ time interval: parameters of the best fits (Band+PL) in the energy ranges (a) $8$--$40000$ keV (GBM) and (b) $8$ keV -- $30$ GeV (GBM+LAT). In the last column of $\Delta T_1$ we list the significance levels from the comparison between nested models (BB+PL over PL, Band+BB over Band and Compt+BB over Compt).}
\label{sign}
\end{table*}

\section{GRB 090510 Theoretical Interpretation}\label{sec3}

In the Fireshell model \citep{Ruffini2001c,Ruffini2001,Ruffini2001a} GRBs originate from an optically thick $e^+e^-$ plasma created by vacuum polarization processes in the gravitational collapse to a black hole \citep{Damour,RVX2010}. 
The dynamics of such an expanding plasma in the optically thick phase is described by its total energy $E^{tot}_{e^+e^-}$ and by the amount of the engulfed baryons $B$.
The spherical symmetry of the system is assumed.
The canonical GRBs light curve is then characterized by a first emission due to the transparency of the $e^+e^-$-photon-baryon plasma, the P-GRB, followed by a multi-wavelength emission, the extended afterglow, due to the collisions, in a fully radiative regime, between the accelerated baryons and the CBM.
The radius at which the transparency occurs, $r_{tr}$, the theoretical temperature blue-shifted toward the observer $kT_{blue}$ and the Lorentz factor $\Gamma_{tr}$ as well as the amount of the energy emitted in the P-GRB are functions of $E_{e^+e^-}^{tot}$ and $B$ \citep{Ruffini2001,Ruffini2009b}.
The structures observed in the extended afterglow of a GRB are described by two quantities associated with the environment: the CBM density profile $n_{CBM}$, which determines the temporal behavior of the light curve, and the filling factor $\mathcal{R}=A_{eff}/A_{vis}$, where $A_{eff}$ is the effective emitting area of the fireshell and $A_{vis}$  its total visible area \citep{Ruffini2002,Ruffini2005}. 
This second parameter takes into account the inhomogeneities in the CBM and its filamentary structure \citep{Ruffini2004}.
The density of each filament is simply defined as $n_{fil}=n_{CBM}/\mathcal{R}$.

We have identified the first episode, where the thermal component is not statistically excluded, with the P-GRB.
Then we have started the simulation using our numerical code \citep[for details, see e.g.][]{Ruffinibrasile}. 
The input parameters are $E_{e^+e^-}^{tot}$, constrained to the isotropic energy of the burst, $E_{iso}=(1.10\pm0.06)\times10^{53}$ erg, and the Baryon load $B=(1.45\pm0.28)\times10^{-3}$, determined by matching the theoretically simulated energy $E_{tr}$ and temperature $kT_{th}=kT_{blue}/(1+z)$ of the P-GRB with the ones observed in the faint pulse, $E_1$ and $kT_{obs}$.
The results of our simulation are the following:
\begin{eqnarray}
\nonumber & \Gamma_{tr} = (6.7\pm1.6)\times10^2 \hspace{0.2cm},\hspace{0.2cm} r_{tr} = (6.51\pm0.92)\times 10^{13}\,\textnormal{cm}\ , \\
\label{val} &E_{tr}=(2.94\pm0.50)\%E_{e^+e^-}^{tot}\hspace{0.2cm},\hspace{0.2cm}kT_{th} = (34.2\pm7.5)\ \textnormal{keV} .
\end{eqnarray}
The theoretically predicted P-GRB energy slightly differs from the observed $E_1=(2.28\pm0.39)\times10^{51}\,\textnormal{erg}=(2.08\pm0.35)\%E_{iso}$, since emission below the threshold is expected between the small precursor and the main emission (see light curves in Fig.~\ref{fig:2}), thus the value of $E_1$ is certainly underestimated.

\begin{table*}
\centering
\begin{tabular}{ccccc}
\hline\hline 
\textbf{Distance}       &  \textbf{$n_{CBM}$}          &  \textbf{$\mathcal{R}$}         &  \textbf{$n_{fil}$}      \\
\textbf{[cm]}           &  \textbf{[$\#$/cm$^3$]}      &                                 &  \textbf{[$\#$/cm$^3$]}  \\
\hline
$6.5\times10^{14}$      &  $550\pm45$                  &  $(3.2\pm0.3)\times10^{-9}$     &  $(1.72\pm0.21)\times10^{11}$ \\
$9.2\times10^{14}$      &  $1.90\pm0.60$               &                                 &  $(5.94\pm0.84)\times10^8$    \\
$1.6\times10^{15}$      &  $60.0\pm4.1$                &                                 &  $(1.88\pm0.22)\times10^{10}$ \\
$2.3\times10^{15}$      &  $(2.50\pm0.20)\times10^3$   &                                 &  $(7.81\pm0.96)\times10^{11}$ \\
$2.5\times10^{15}$      &  $0.15\pm0.01$               &                                 &  $(4.69\pm0.53)\times10^{7}$  \\
\hline
$3.3\times10^{15}$      &  $(1.90\pm0.20)\times10^4$   &  $(1.5\pm0.2)\times10^{-10}$    &  $(1.27\pm0.22)\times10^{14}$ \\
$3.4\times10^{15}$      &  $0.15\pm0.02$               &                                 &  $(1.00\pm0.19)\times10^{9}$  \\
\hline
$3.5\times10^{15}$      &  $(2.50\pm0.14)\times10^4$   &  $(3.8\pm0.4)\times10^{-8}$     &  $(6.58\pm0.78)\times10^{11}$ \\
$3.6\times10^{15}$      &  $0.10\pm0.02$               &                                 &  $(2.63\pm0.59)\times10^6$    \\
\hline
\end{tabular}
\caption{We report for each cloud, respectively, the distance from the black hole, the average number density (assuming spherically distributed clouds), the filling factor and the number density of the filaments.}
\label{cbm}
\end{table*}

\begin{figure}
\centering
\includegraphics[width=0.9\hsize,clip]{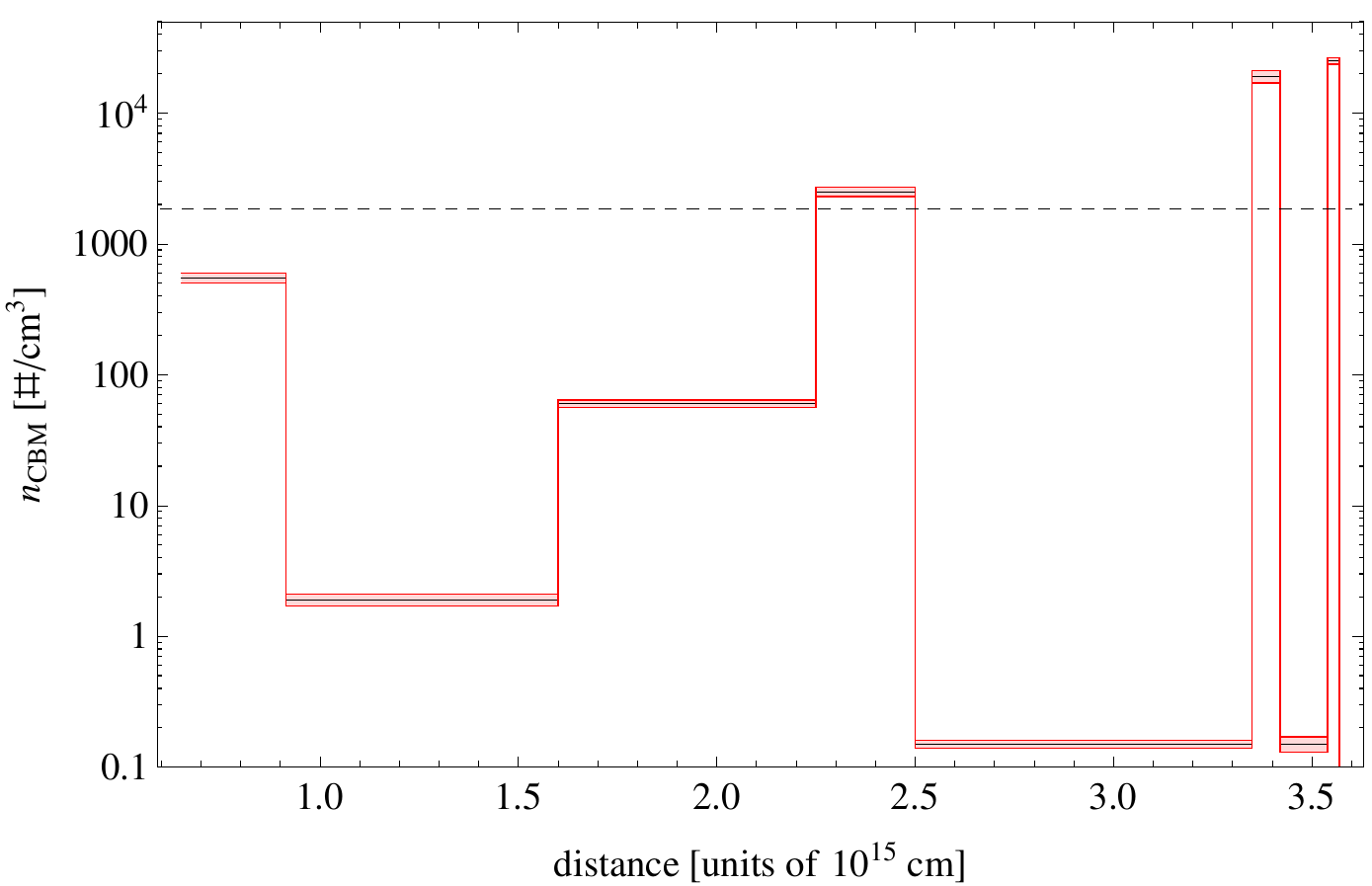}\\
\includegraphics[width=0.9\hsize,clip]{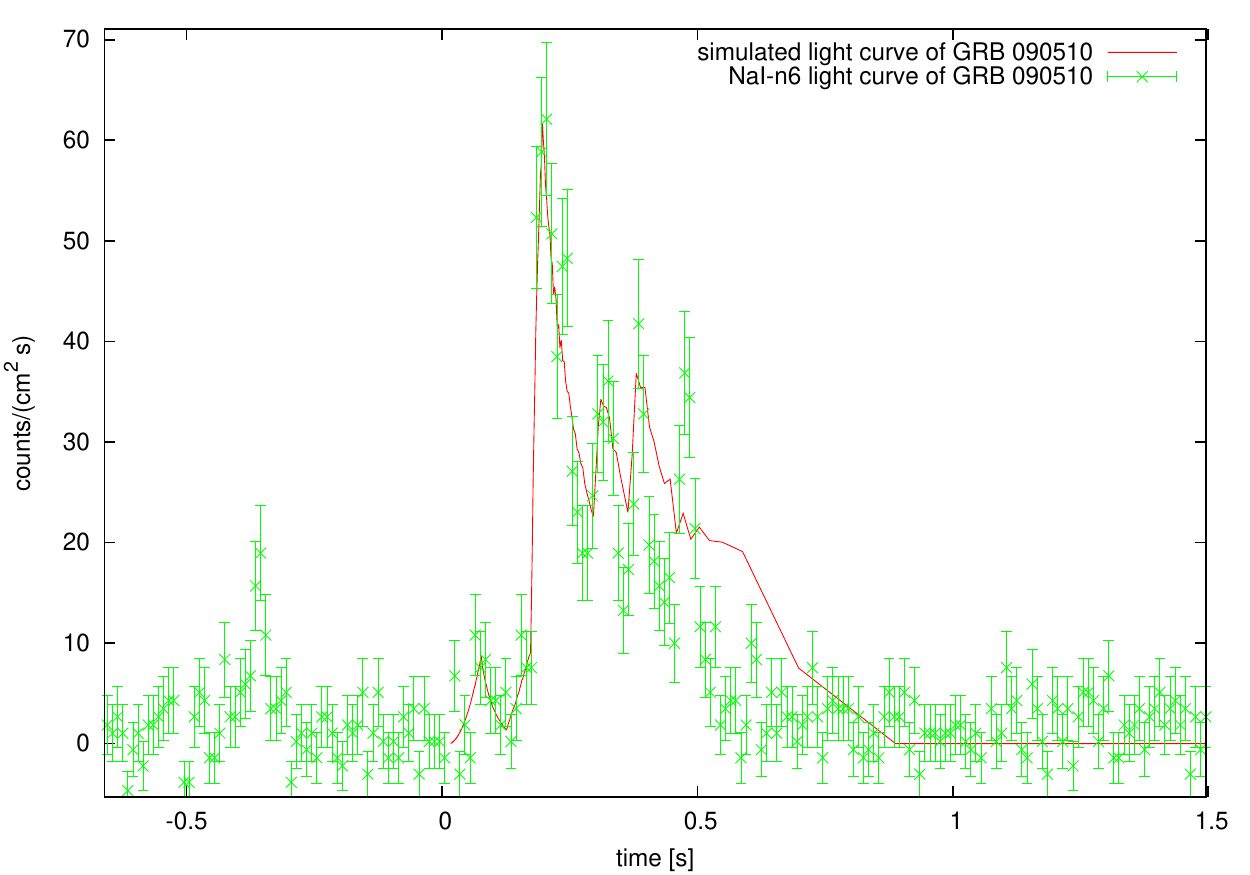}\\
\includegraphics[width=0.9\hsize,clip]{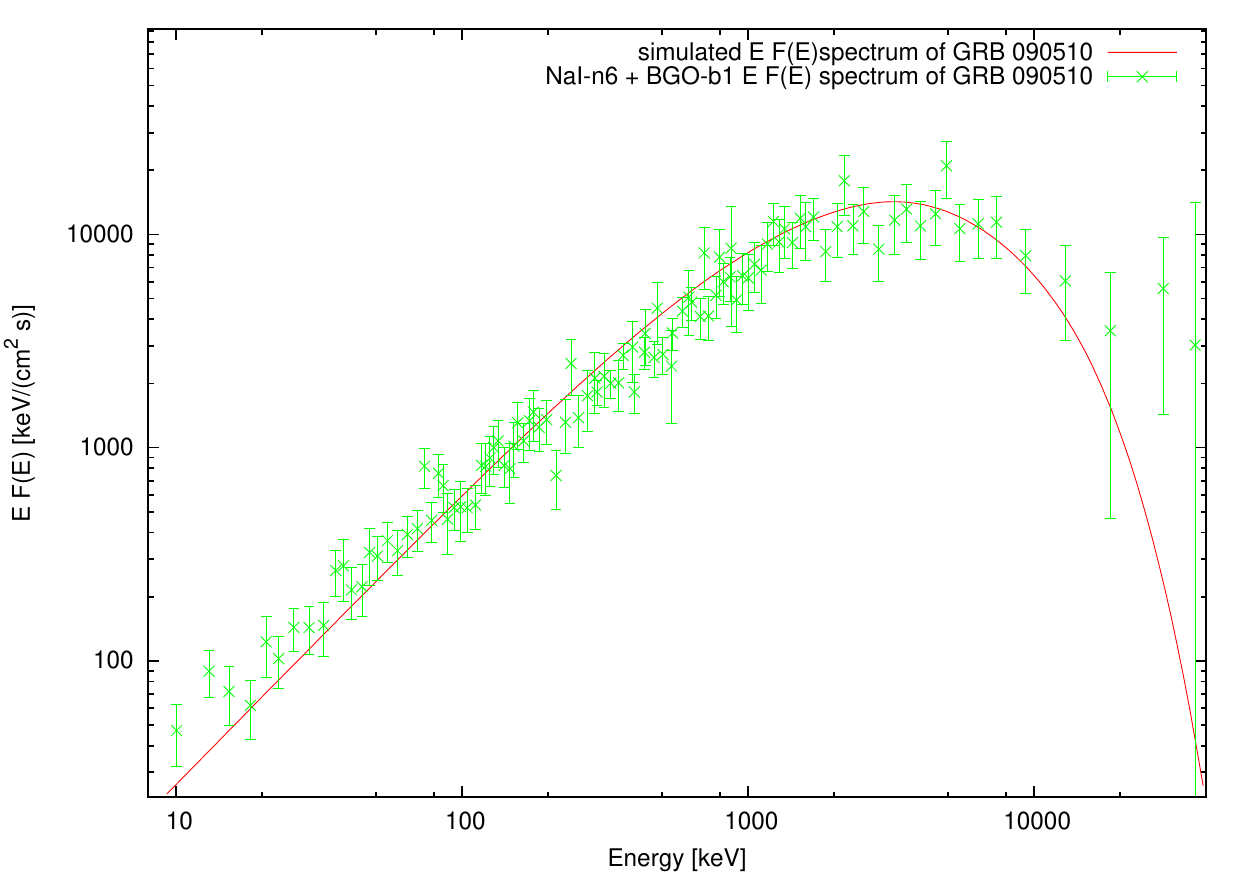}
\caption{In the upper panel the radial CBM density distribution of GRB 090510 (red solid line) with its uncertainty (light red shaded region) and the mean value (black dashed line) are shown. The simulated NaI-n6 light curve ($8$--$1000$ keV) of the extended afterglow (middle panel) and the corresponding spectrum of the early $\sim0.4$ s of the emission in the energy range $8$ keV -- $40$ MeV (lower panel) are consequently obtained. }
\label{fig:3}
\end{figure}

In the following analysis we focus our attention on the main emission.
Since in $\Delta T_2$ no evidence of a thermal component has been found (see Fig.~\ref{fig:2}, bottom right panel, and table below), we have interpreted this emission as the extended afterglow. 
Using the above values of $E_{e^+e^-}^{tot}$ and $B$, we have simulated the light curve of the extended afterglow by defining the radial number density distribution of the CBM (assuming spherically distributed clouds) and the value of the filling factor $\mathcal{R}$, following a trial and error procedure to reproduce the pulses observed in the light curve and the corresponding spectrum.
The errors on the densities and the filling factors are obtained varying them within the observational errors; typically the errors are about 10\% of the value.
The average value is indeed very high, $\langle n_{CBM} \rangle=(1.85\pm0.14)\times10^3$ particles/cm$^3$, assuming spherically distributed clouds (see Fig.~\ref{fig:3}, upper plot).
Basically this high average density is due to the second and the third brightest spikes of the light curve (see Fig.~\ref{fig:3}, middle panel), where the density of the clouds is $\sim2\times10^4$ particles/cm$^3$ (see Tab.~\ref{cbm}, second column).
The filling factor assumes values $1.5\times10^{-10}\leq\mathcal{R}\leq3.8\times10^{-8}$ (see Tab.~\ref{cbm}, third column). 
Correspondingly, the values of the densities of the filaments $n_{fil}$ are estimated (see Tab.~\ref{cbm}, fourth column).
In Fig.~\ref{fig:3} we show also the simulated extended afterglow light curve from the NaI-n6 detector (middle panel) and the corresponding spectrum of the early $\sim0.4$ s of the emission (lower panel) in the energy range $8$ keV -- $40$ MeV, using the spectral model described in \citet{Bianco2004} and \citet{Patricelli2012}.
The last part of the simulation requires a more detailed $3$-dimensional code to take into due account the distribution of the CBM.

\section{Conclusions}\label{sec4}

We list our conclusions.

1) The simulated spectrum of the extended afterglow in the time interval $\Delta T_2$, considered in the analysis by \citet{Ackermann2010}, is in excellent agreement with the one in Fig.~\ref{fig:2} in the sub-MeV and in the MeV region.
The baryon load $B=(1.45\pm0.28)\times10^{-3}$ used in this simulation has been determined from the analysis of the first episode, which has been identified with the P-GRB.
The current quality of the data does not allow us to properly distinguish between BB+PL and Compt spectral models.
From our theoretical interpretation, BB+PL model was adopted, since it is not ruled out by the data.
Such a fundamental issue will be further clarified by future space missions with larger collecting area and X/$\gamma$-rays timing, as e.g. the LOFT mission \citep{LOFT2012}.

2) We have stressed a key difference between the Fireshell and the Fireball approaches.
In the Fireshell model the extended afterglow encompasses the prompt emission and the afterglow of the traditional Fireball model.
The density of the CBM is inferred from the prompt emission by assuming the fully radiative condition emission in a optically thin regime \citep{Ruffini2002,Ruffini2004,Ruffini2005} and a precise spectrum in the comoving frame is assumed \citep{Patricelli2012} and convoluted over the EquiTemporal Surfaces \citep[EQTS,][]{Bianco2005b,Bianco2005a}.
In the Fireball model, instead, the density is estimated from the afterglow emission by analyzing emission or absorption lines in the X-ray spectra \cite[see e.g.][]{Lazzati1999,Prochaska2008}, or by observing steepening or breaks of the optical afterglow light curves \cite[see e.g.][]{Dai1999,Piro2001,Wang2003}. 
From the fully radiative condition, we have found that GRB 090510 occurs in an over-dense medium with an average value of $\langle n_{CBM} \rangle\approx10^3$ particles/cm$^3$ (for spherically symmetric distributed clouds).
This high CBM density and the small value of the filling factor, $1.5\times10^{-10}\leq\mathcal{R}\leq3.8\times10^{-8}$, leads to local over-dense CBM clouds, in the form of filaments, bubbles and clumps, with a range of densities $n_{fil}=n_{CBM}/\mathcal{R}\approx(10^{6}-10^{14})$ particles/cm$^3$.

3) The joint effect of the high value of the Lorentz factor, $\Gamma_{tr}=(6.7\pm1.6)\times10^2$, and the high density compresses in time the emission of the extended afterglow. 
Therefore its light curve is shortened in time and ``inflated'' in intensity with respect to the canonical one for disguised short bursts (see Fig.~\ref{fig:1}, lower panel), making it apparently closer to the genuine short class of GRBs \citep{Muccino2012}.
It is interesting to note that in this GRB, with an abnormally high value of the CBM density, the extended afterglow does not fulfill the Amati relation \citep{Amati2006}.
%In \citet{Izzo2012b} it has been inferred a fireshell with a Lorentz factor at the transparency $\Gamma_0=142\pm57$ traveling in a CBM with average density $\langle n_{CBM} \rangle=3.4\times10^3$ particles/cm$^3$ (for spherically symmetric distributed clouds) and giving rise to an extended afterglow emission lasting $\Delta t_0\sim40$ s. 
%In the case of GRB 090510, the CBM average density has the same order of magnitude as for GRB 970828, but the Lorentz factor at the transparency is $\sim6$ times greater.
%Therefore, taking into account the Lorentz transformations from the laboratory frame to the local observer one and assuming for the sake of simplicity a constant Lorentz factor all over the extended afterglow emission, the duration of GRB 090510 should be roughly $\Delta t=\Delta t_0(\Gamma_0/\Gamma_{tr})^2\approx1.8$ s.
%This result, as a first order approximation, is consistent with the $T_{90}$ of GRB 090510.}

%\textbf{typically associated with the star forming regions with dense molecular clouds \citep{Piro2001}.
%As discussed in literature \citep{Dai1999,Piro2001,Wang2003}, high density environments with $n\sim10^4$--$10^6$ particles/cm$^3$ show the presence of breaks in the their optical or near infrared afterglow occurring around one day after the trigger.
%In the case of GRB 090510, because of the densities in some filaments up to $10^{14}$ particles/cm$^3$, this break may occur in the X-ray afterglow and because of the high value of $\Gamma_{tr}$ it can occur in less than one day after the trigger.}

4) From the values of $n_{fil}$ we obtain a range of grammages of $m_H n_{fil} \Delta r_c\approx(10^{-2}-10^{4})$ g/cm$^2$, where $m_H$ is the mass of the Hydrogen atom and $\Delta r_c$ is the size of the cloud inferred from our simulation (see Fig.~\ref{fig:3} and the first column in Tab.~\ref{cbm}).
This high value of the grammage may be relevant in the explanation of the observed GeV emission as originating in the collisions between ultra high energy protons, with bulk Lorentz factor of $\Gamma_{tr}=(6.7\pm1.6)\times10^2$, and the CBM.

\end{document}